\documentclass[final,5p,times,twocolumn]{elsarticle}

\usepackage{graphicx}
\usepackage{epsfig}
\usepackage{amssymb}
\usepackage{lineno}
\usepackage[percent]{overpic}
\usepackage{natbib}

\journal{Chemical Physics Letters}

\begin{document}

\begin{frontmatter}
\title{Equality of diffusion-limited chronoamperometric currents to equal area spherical and cubic nanoparticles on a supporting electrode surface}

\author[Oxford]{Enno K\"atelh\"on}
\author[Oxford]{Edward O. Barnes}
\author[Juelich]{Kay J. Krause}
\author[Juelich,Aachen]{Bernhard Wolfrum}
\author[Oxford]{Richard G. Compton\corref{cor1}}

\ead{richard.compton@chem.ox.ac.uk}
\cortext[cor1]{Corresponding author. Fax: +44 (0) 1865 275410}

\address[Oxford]{Department of Chemistry, Physical and Theoretical Chemistry Laboratory, Oxford University, South Parks Road, Oxford OX1 3QZ, United Kingdom}
\address[Juelich]{Institute of Bioelectronics (PGI-8/ICS-8) and JARA-Fundamentals of Future Information Technology, Forschungszentrum J\"ulich, 52425 J\"ulich, Germany}
\address[Aachen]{IV. Institute of Physics, RWTH Aachen University, 52074 Aachen, Germany}

\begin{abstract}
We computationally investigate the chronoamperometric current response of spherical and cubic particles on a supporting insulating surface. By using the method of finite differences and random walk simulations, we can show that both systems exhibit identical responses on all time scales if their exposed surface areas are equal. This result enables a simple and computationally efficient method to treat certain spherical geometries in random walk based noise investigations.
\end{abstract}

\end{frontmatter}

\section{Introduction}

Random walk simulations provide a versatile tool for a variety of applications in electrochemistry \citep{licht_time_1989, nagy_three-dimensional_1997, nagy_three-dimensional_1997-1, white_random_2005, singh_stochasticity_2012}. Recent advances in computational power and the wide availability of multi-core systems \cite{katelhon_noise_2013} and massively parallelised graphics processing units (GPUs) \cite{cutress_how_2011, cutress_electrochemical_2011} opened up new options for high-performance simulations involving large numbers of molecules as well as a high resolution. Hereby, random walks can particularly be employed to model mesoscopic effects as well as noise characteristics of nanoscale sensors that cannot be simulated through numerical methods based on solutions of differential equations. 

However, even though random walks are successfully applied in many studies, their use may still remain challenging in certain geometries. Difficulties mostly arise from the general concept of the random walk: in random walk simulations, molecules are typically displaced by a random vector of a fixed length $dr$ at a frequency of $1/dt$, where $dr$ and $dt$ fulfil

\begin{equation}
	dr = \sqrt{2nDdt}
\end{equation}
in the $n$-dimensional case, where $D$ represents the diffusion coefficient. However, this equation is a direct consequence of the general solution of the diffusion equation, which is only valid in absence of physical boundaries. While this can be compensated in the case of flat surfaces \cite{katelhon_noise_2013}, rough or porous surfaces can only be modelled provided $dr$ is significantly smaller than the investigated structure and appropriate boundary conditions have been applied. Furthermore, surfaces that are curved in regard to the displacement vector's coordinate system are difficult to model, since boundary conditions have to be separately defined for every single grid point. This is, for instance, the case for the later discussed geometry of a spherical electrode in a Cartesian coordinate system.

An alternative approach to electrochemical simulations is given by the method of finite differences, which has been used in a wide range of applications during recent years \cite{melville_hydrodynamics_2003, britz_digital_2005, britz_comparison_2008, amatore_theory_2008, amatore_new_2010, barnes_electrochemical_2011}. In comparison to random walks, it is a relatively fast method for the investigation of electrochemical systems. However, since the locations of single molecules are described through probability density functions that spread across the whole simulated space, stochastic processes in the mesoscopic regime and, hence, noise characteristics of electrochemical sensors cannot be analysed.

Here, we address the modelling of nanosphere electrodes through random walks. Investigated spheres are mounted on an insulating flat surface, which is a typical experimental setting for electrodes that are modified with nanoparticles or certain nanoimpact setups. While previous attempts at equivalence problems focused on the comparison between microdisc- and hemispherical electrodes \cite{alden_can_1997} and demonstrated that in this case equal 'superficial radii' lead to equal steady-state voltammograms \cite{oldham_comparison_1988}, we here find an equivalent for the direct modelling of the sphere electrode through the random walks in Cartesian coordinates. We first calculate an exact solution for a single sphere via finite differences before we fit this result to cubic electrodes that were modelled through random walks. By this method, we find an equivalent cube that features an identical chronoamperometric current response. The presented equality holds true on all time scales and is independent of the electrode size. Hence, all findings are equally applicable to microparticles or even larger particles.

\section{Theoretical model}
The model considered is a simple one-electron reduction process,

\begin{equation}
  A + e^- \rightarrow B
\end{equation}
where molecules are assumed to react immediately and fully upon contact with the electrode, that is to say the process is diffusion controlled. An excess amount of supporting electrolyte is also considered to be present in solution, supressing electric fields and ensuring that mass transport is limited to diffusive processes. Hence, electrode currents are solely limited by the analyte's diffusive mass transport towards the electrode surface, which is a good approximation for many diffusion-limited chronoamperometric experiments.

In this work, we investigate two electrode geometries: a spherical and a cubic electrode on an insulating surface, as shown in Figure 1. For the sphere on an insulating surface, cyclic voltammetry has previously been simulated \cite{ward_modelling_2012} using a Fickian diffusion model and the finite difference method, which is here modified to simulate chronoamperometry. A stochastic random walk model is used to simulate chronoamperometry at a cubic electrode, and the corresponsence between the two cases is investigated. Simulation and computational procedures are detailed below.

\begin{figure}[t]
  	\centering
	\begin{overpic}[width=0.5\textwidth]{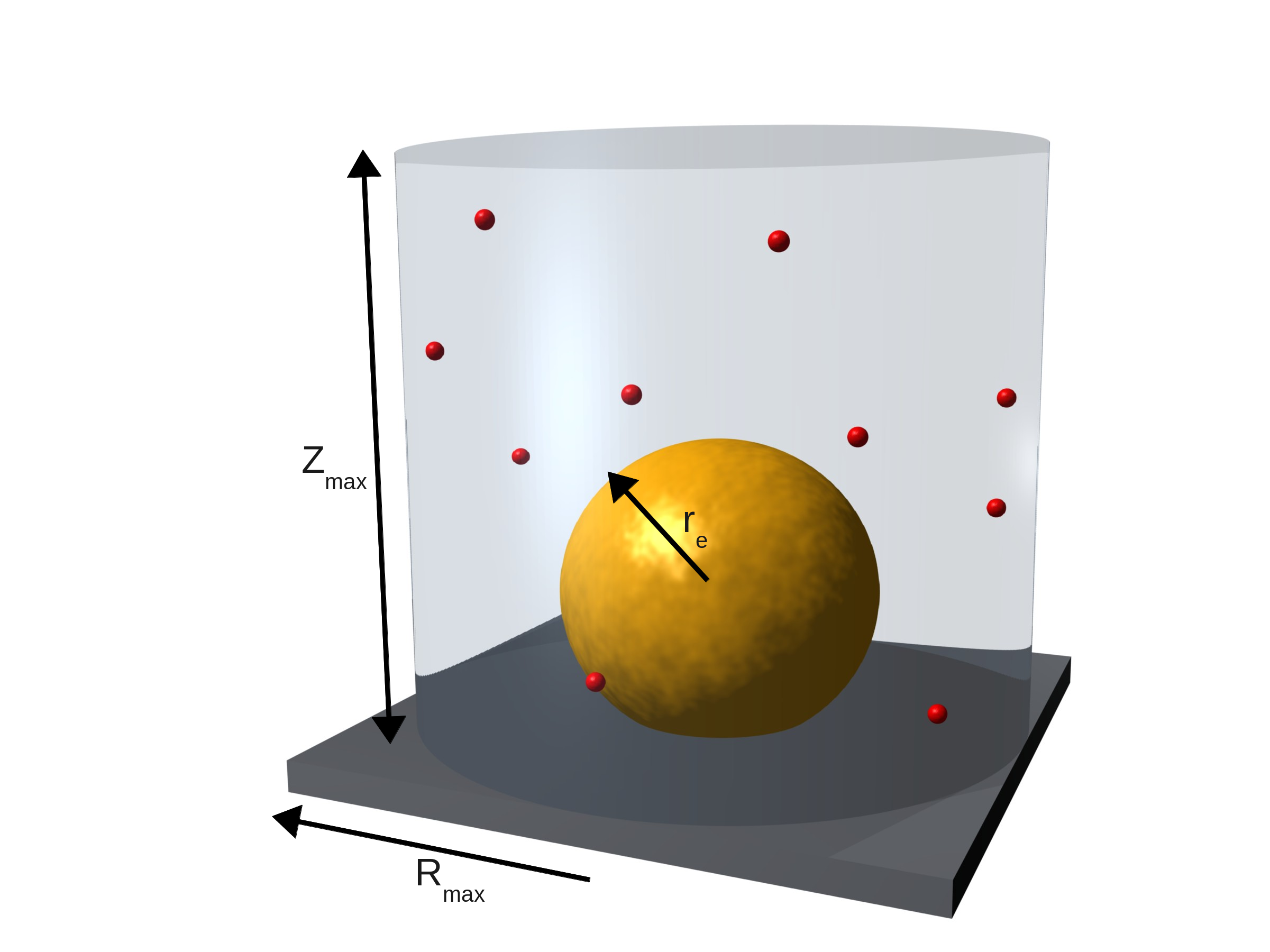}
		\put (0,65) {a)}
	\end{overpic}
	\begin{overpic}[width=0.5\textwidth]{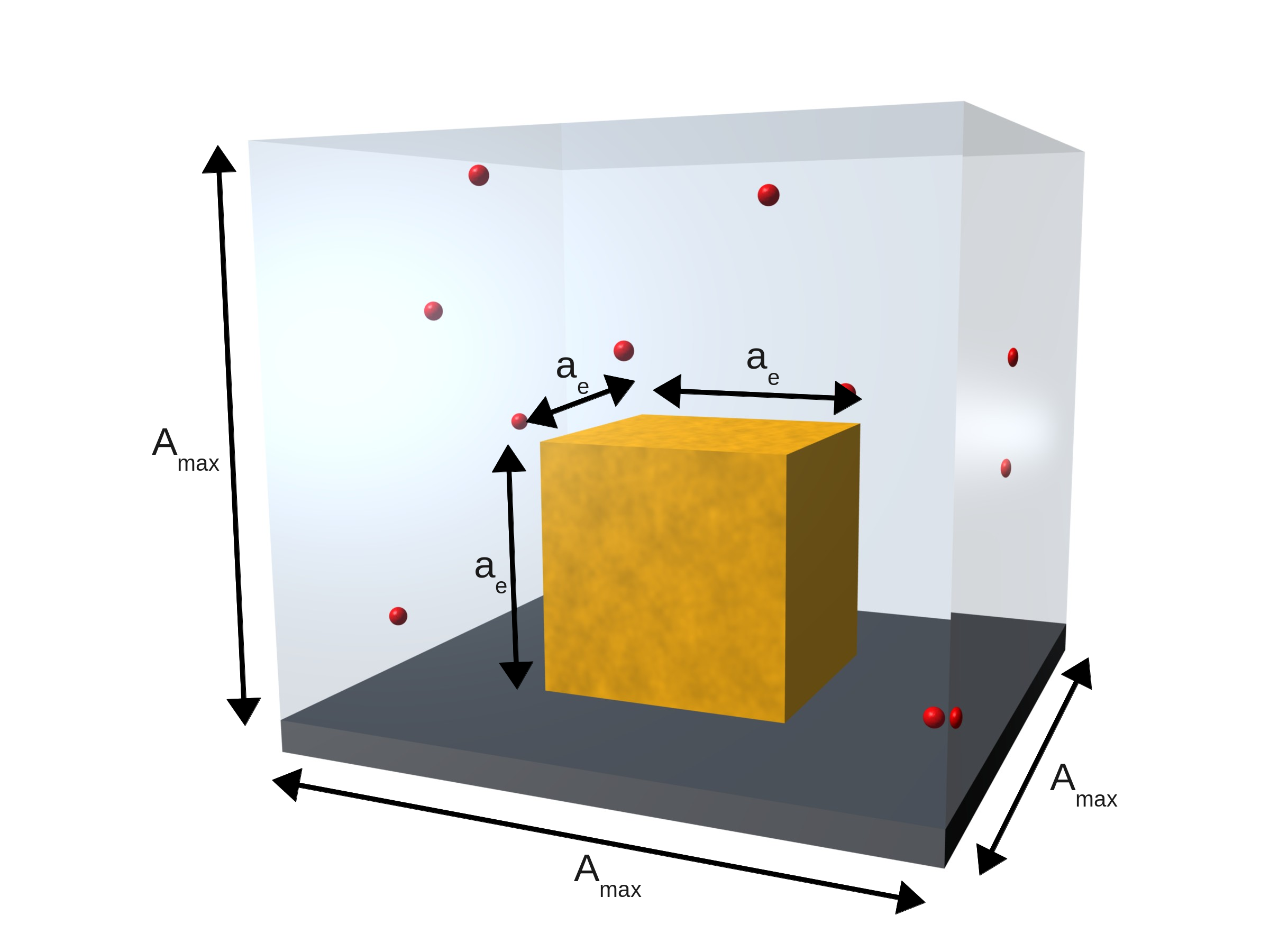}
		\put (0,65) {b)}
	\end{overpic}  	
	\caption{Schematic of the two simulated geometries: a spherical (a) and a cubic (b) electrode on a supporting, electrochemically inactive surface. Dimensions are not to scale.}
\end{figure}

\subsection{Finite differences}
In order to simulate electrochemistry at an isolated spherical electrode resting on an insulating surface, symmetry can be exploited to reduce the problem to two dimensions. A schematic illustration of this is shown in Figure 2.

\begin{figure}[t]
\centering
\includegraphics[width=0.2\textwidth]{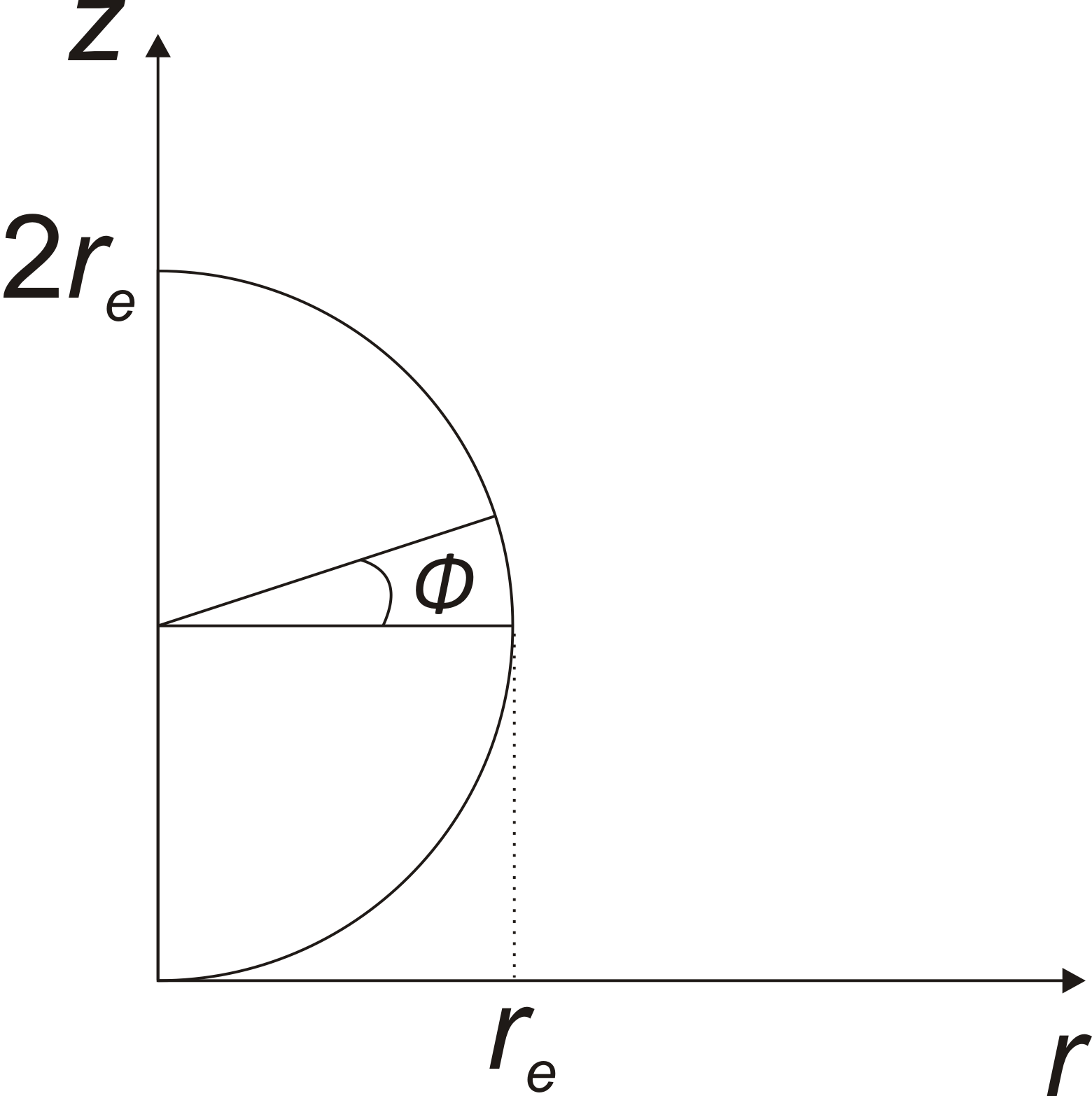}
\caption{Schematic diagram of the simulation space used to simulate an isolated spherical electrode on an insulating surface.}
\label{ELECTRODE SCHEME}
\end{figure}

Fick's second law for diffusion in cylindrical space is given by:

\begin{equation}
	\frac{\partial{c}}{\partial{t}} = D\left(\frac{\partial^2{c}}{\partial{r^2}} + \frac{\partial^2{c}}{\partial{z^2}} + \frac{1}{r}\frac{\partial{c}}{\partial{r}}\right)
\end{equation}

To simplify the model, dimensionless parameters are used in simulations. Concentrations are taken relative to the bulk concentration of the electroactive species ($c^*$), and the relevant diffusion coefficient taken to be unity. Dimensionless parameters are defined in Table \ref{DIMENSIONLESS}.

\begin{table}
\begin{center}
\begin{tabular}{l l}
\hline
Normalised Parameter & Definition\\
\hline
$C$ & $\frac{c}{c^*}$\\
$\tau$ & $\frac{D}{r_e^2}t$\\
$R$ & $\frac{r}{r_e}$\\
$Z$ & $\frac{z}{r_e}$\\
\hline
\end{tabular}
\end{center}
\caption{Normalised parameter definitions}
\label{DIMENSIONLESS}
\end{table}

Substituting these parameters into the mass transport equation, the following equation is obtained:

\begin{equation}
	\frac{\partial{C}}{\partial{\tau}} = \left(\frac{\partial^2{C}}{\partial{R^2}} + \frac{\partial^2{C}}{\partial{Z^2}} + \frac{1}{R}\frac{\partial{C}}{\partial{R}}\right)
\end{equation}
This must be solved over all space subject to appropriate boundary conditions. After the beginning of the experiment at $\tau=0$, the electrode is biased to a potential to oxidise the electroactive species at a mass transport controlled rate. The concentration at the electrode surface can therefore be defined as zero:

\begin{equation}
	C_{R^2 + \left(Z-1\right)^2 = 1} = 0
\end{equation}

The outer edges of the simulation space are set to be well outside the depletion layer around the electrode:

\begin{eqnarray}
	R_\mathrm{max} = 1 + 6\sqrt{\tau_\mathrm{max}}\\
	Z_\mathrm{max} = 2 + 6\sqrt{\tau_\mathrm{max}}
\end{eqnarray}
where $\tau_\mathrm{max}$ is the total dimensionless time of the experiment \cite{gavaghan_exponentially_1998_1, gavaghan_exponentially_1998_2}. At these boundaries, the (dimensionless) concentration can be set equal to unity. The flux of species across the insulating surface which the electrode is resting on and the axisymmetric $Z$ axis can set to be zero:

\begin{equation}
	\left(\frac{\partial{C}}{\partial{Z}}\right)_{Z=0} = \left(\frac{\partial{C}}{\partial{R}}\right)_{R=0,Z>2} = 0
\end{equation}

There is now a complete set of equations to be solved simultaneously over all space at each point in time to obtain a concentration profile. These equations are descretised using the Crank-Nicolson method \cite{crank_practical_1996} to a grid of spatial \cite{streeter_diffusion-limited_2007} and temporal \cite{barnes_dual_2013} points. The two dimensional problem is solved using the alternating direction implicit (ADI) method. The Thomas algorithm \cite{press_numerical_2007} is used to solve the resulting tri-diagonal matrices. The simulations were coded in C++ and run on an Intel (R) Xeon (R) 2.26 GHz PC with 2.25 GB RAM, with a typical run time of approximately 20 minutes.

\subsubsection{Calculation of the current}

The total flux over the electrode can be found by integrating the flux density over the whole electrode surface. The (dimensionless) flux density at any point is given by:

\begin{equation}
	j = \frac{\partial{C}}{\partial{N}}
\end{equation}
where $N$ is some coordinate normal to the surface of the electrode at that point. In terms of $R$ and $Z$, the flux density is calculated as:

\begin{equation}
	j_\phi = \frac{\partial{C}}{\partial{R}}\mathrm{cos}\phi + \frac{\partial{C}}{\partial{Z}}\mathrm{sin}\phi
\end{equation}

The total (real) current can then be obtained as follows assuming a single electron is transferred as in Equation (2):

\begin{equation}
	I = 2\pi F\int^{\pi/2}_{-\pi/2}{j_\phi \mathrm{cos}\phi}\mathrm{d}\phi
\end{equation}

\subsection{Random walks}
Random walk simulations are performed using a previously reported simulation framework \cite{katelhon_noise_2013}. In short, active molecule trajectories are modelled individually as random walks with each dimension is treated independently. Upon collision with simulation boundaries, pathways are reflected while contact with the electrode causes an immediate charge transfer. The simulation is programmed in C/C++ and employs the application programming interface Open Multi-Processing (OpenMP) for parallel computing. All further data analysis is done in Matlab.

In this work, all calculations are based on a diffusion coefficient of 9.3$\times$10$^{-9}$ m$^2$ s$^{-1}$, which equals the diffusion coefficient of protons in water \cite{CRC}. We further set $A_{max}$ to 500 nm and calculate trajectories of 5x10$^{6}$ individual moleculues per simulation, which corresponds to a bulk concentration of 66.4 mM. The outer boundaries of the simulation space are hereby chosen to extend far beyond the diffusion field of the experiment and fulfil:

\begin{equation}
  \frac{A_{max} - a_e}{2} > 6 \sqrt{Dt_{max}}
\end{equation}
where $t_{max}$ represents the duration of the experiment.

\section{Results and discussion}
In this section, we compare the dimensionless solution of the chronoamperometric current of a spherical electrode to random walk simulations of cubic electrodes of different sizes. We vary the one-dimensional step width $dr$ of the random walk and investigate its impact on the obtained results.

In order to successfully model a spherical electrode as a cube in a random walk simulation, we can vary the dimensions of each to see, for example, what radius of a sphere is needed to produce a chronoamperogram which fits that produced at a cube of the side length $a_e$. Since the diffusion coefficient and the bulk concentration remain equal in both cases, we can vary the sphere's radius to determine the best fit $r_{fit}$, while all other parameters remain constant. Therefore, we solve the following optimization problem:

\begin{equation}
	r_{fit}(a_e) = \min\limits_{r \in \mathbb{R}}  \int dt \left(I_{fd}(t,r) - I_{rw}(t,a_e)\right)^2
\end{equation}
where $I_{fd}$ and $I_{rw}$ represent the currents that are calculated via the finite differences and the random walk simulations. Examples of results can be found in Figure 3, in which we additionally plot the current of a sphere that features a surface area equal to the exposed surface of the cube electrode for later comparison. The radius $r_{es}$  of this sphere (`es' abbreviates `equal surface') can be calculated to be:

\begin{equation}
	\label{cube_sphere_relation}
	r_{es} = \sqrt{ \frac{5}{4 \pi}} a_e
\end{equation}

\begin{figure}[t]
  	\centering
	\begin{overpic}[width=0.5\textwidth]{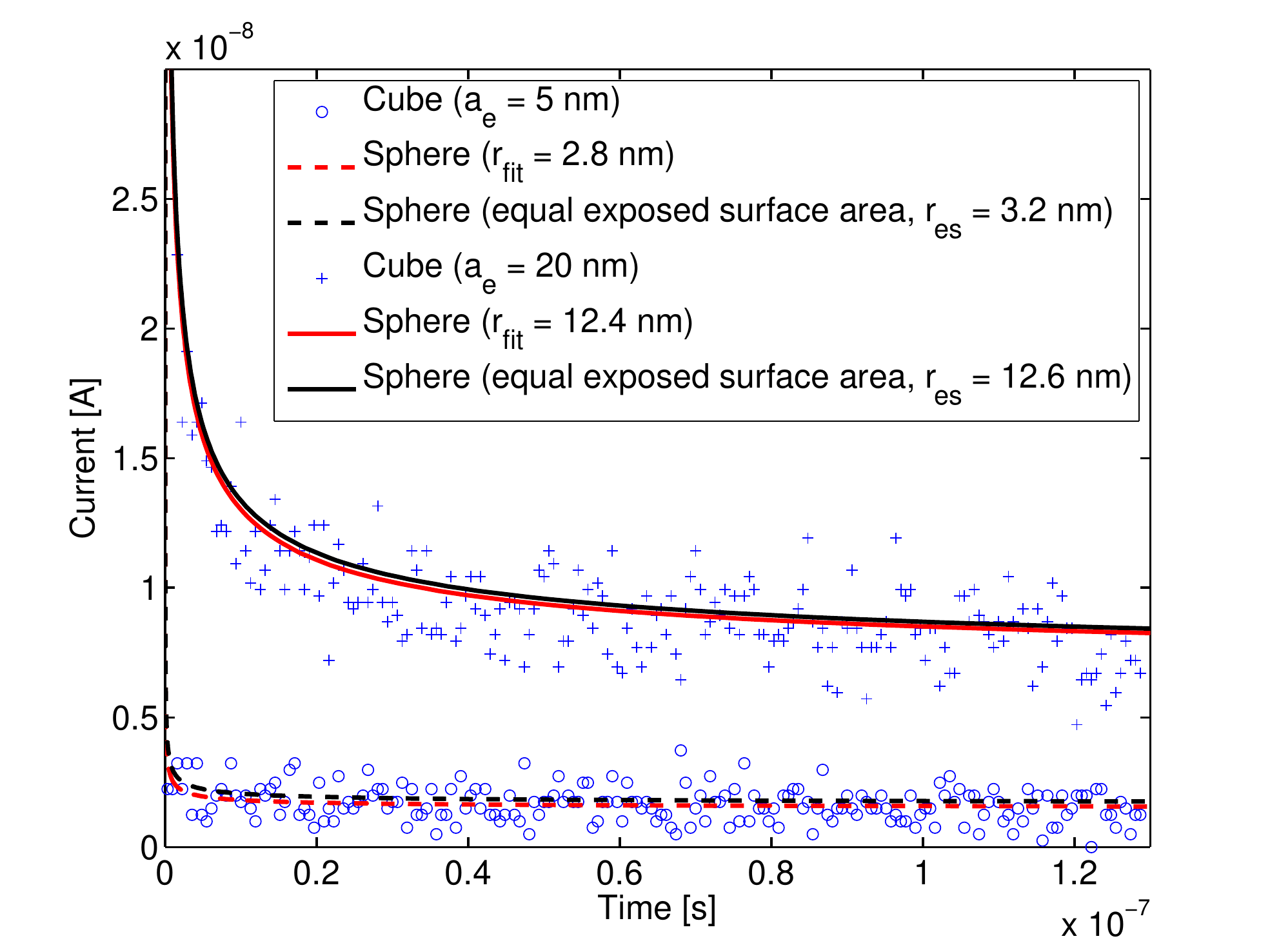}
		\put (0,70) {a)}
	\end{overpic}
  	\begin{overpic}[width=0.5\textwidth]{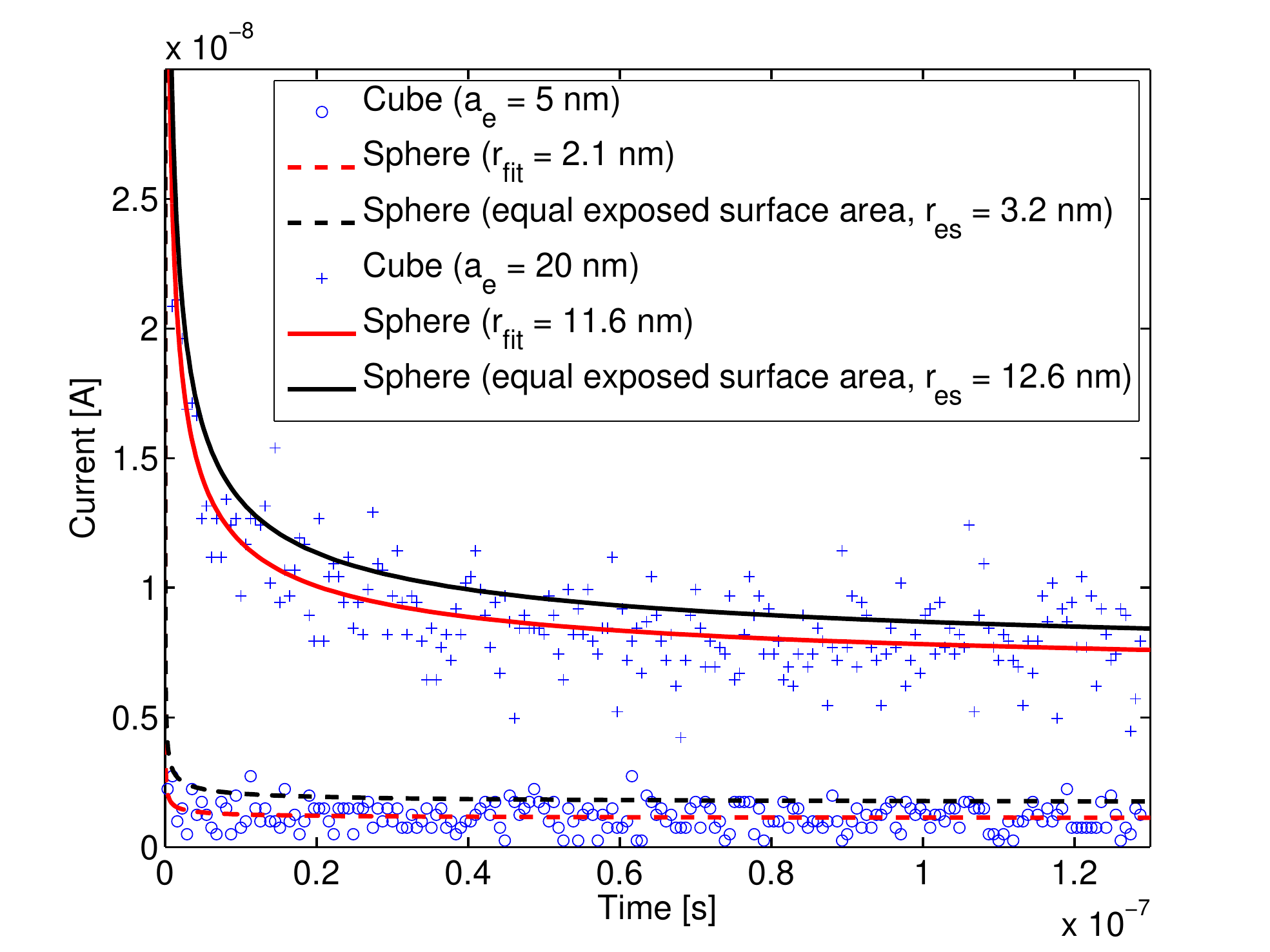}
		\put (0,70) {b)}
	\end{overpic}
  	\caption{Simulated current responses in chronoamperometry. Both plots present the currents of a 5 and a 20 nm cube in comparison to spheres featuring either the fitted radius $r_{fit}$ or a surface area that equals the exposed surface area of the cube. (a) and (b) show data obtained at a random walk step width of 0.25 and 1 nm, respectively.}
\end{figure}

If we compare the graphs in Figure 3a and 3b, we can find that the currents obtained through finite differences simulations and random walks match each other well on all time scales. While in the short term limit electrodes featuring equal surface areas will always exhibit equal chronoamperometric responses, this does not necessarily hold true for all time regimes. Nevertheless, the here presented equality criteria also exhibits good applicability in the long term limit and in intermediate regions, as can be seen in case of the 5 nm and the 20 nm cube, respectively. However, even though both models match each other well, the magnitude of the current as well as the radius that was calculated in the fit decrease with an increasing $dr$. Furthermore, we can find that $r_{fit}$ approaches the value of $r_{es}$ at decreasing $dr$, which will be analysed in more detail below. This dependency of the result on the chosen step width can be understood through the formerly mentioned shortcomings of the random walk approach. Since the random walk reproduces the general solution of the diffusion equation, its results are only accurate in free space while edges and corners are not modelled exactly. By varying the step width relative to the size of the cube, we also vary the overall space occupied by edges and corners. Hence, one can expect an increase in accuracy when moving to smaller, yet computationally less efficient, step widths.

Figure 4a compares the results of $r_{fit}$ to $r_{es}$, where we again find that the value of $r_{fit}$ approaches $r_{es}$ with a decrease in $dr$. Furthermore, the plot shows a step-width dependent offset that is independent from the cube's side length. This result supports our former interpretation of the deviations between different random walk simulations that the underestimation of the current appears to be mostly caused by the inaccuracy of the random walk approach at edges and corners. However, it also suggests to compare the obtained results in terms of the ratio between the $a_e$ and $dr$ as it is shown in Figure 4b. This figure demonstrates that smaller $dr$ leads to smaller deviations between $r_{fit}$ and $r_{es}$, while $r_{fit}$ matches $r_{es}$ in the limit $dr \rightarrow 0$. This results allows the conclusion that spherical electrodes on an inactive surface can be modelled through cubic electrodes as long as they feature equal exposed surface areas, i.e. their dimensions fulfil Equation (\ref{cube_sphere_relation}).

\begin{figure}[t]
  	\centering
	\begin{overpic}[width=0.5\textwidth]{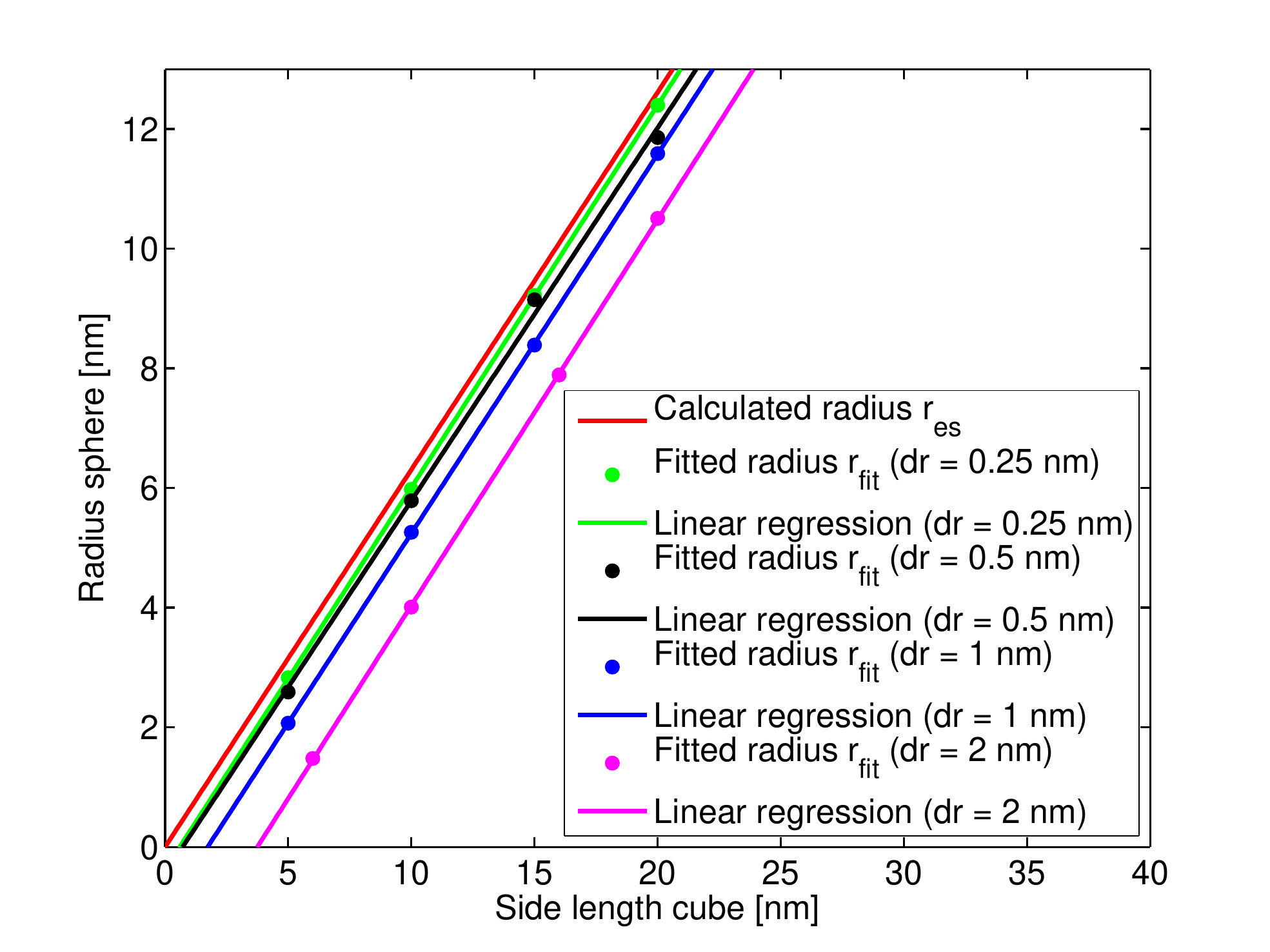}
		\put (0,70) {a)}
	\end{overpic}
  	\begin{overpic}[width=0.5\textwidth]{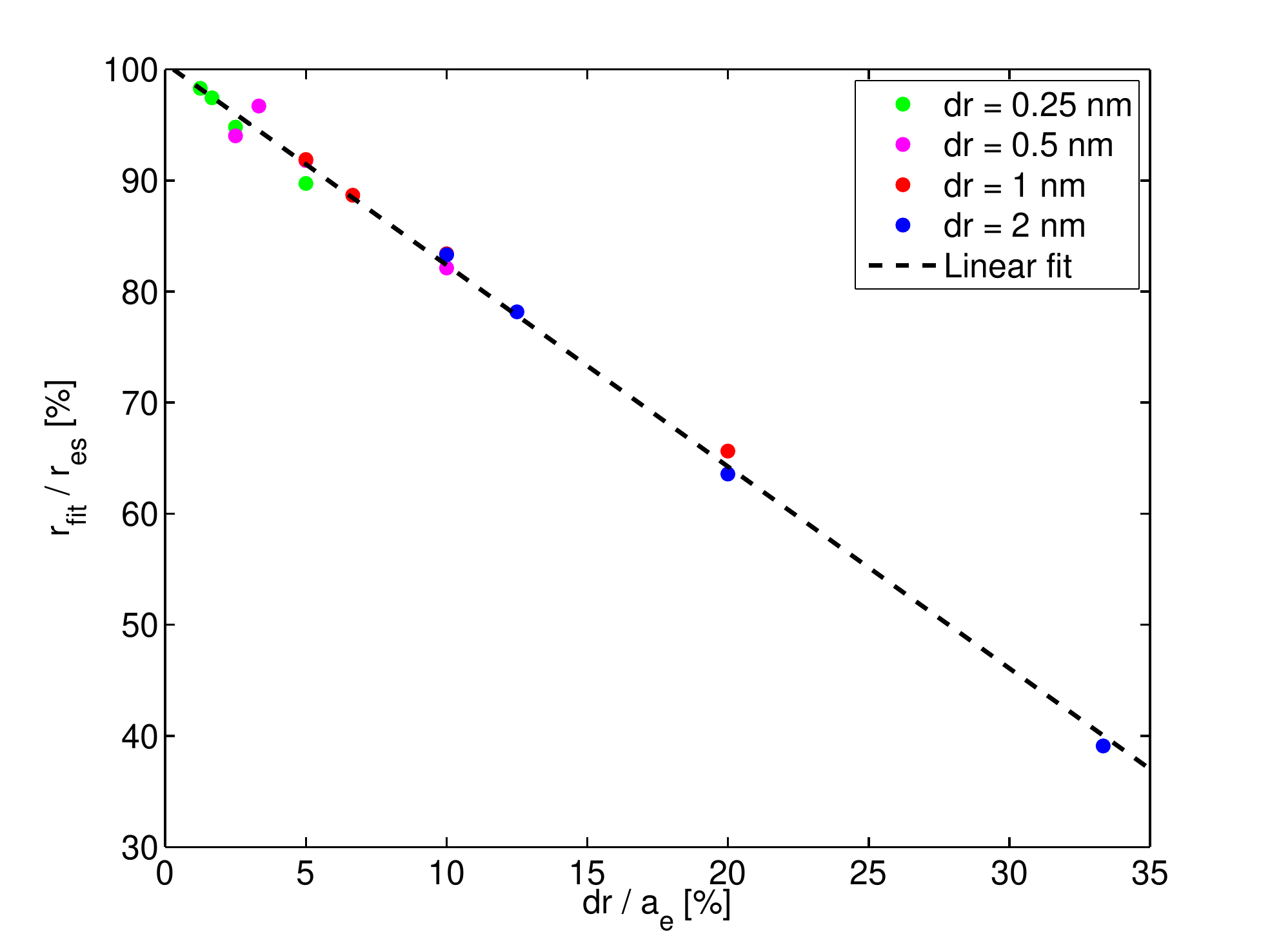}
		\put (0,70) {b)}
	\end{overpic}
  	\caption{Comparison between the $r_{es}$ and the computationally determined values of $r_{fit}$ for different random walk step widths $dr$. (a) provides absolute values, while (b) sets $dr$ and $a$ in relation to each other.}
\end{figure}

\section{Conclusions}
In this work, we compare the chronoamperometric response of spherical and cubic electrodes on a supporting, electrochemically-inactive electrode. While using a finite differences approach for the spherical- and random walks for the cubic geometry, we can show that both geometries lead to equal current responses on all time scales provided the exposed surface areas equal each other.

For future studies, this result enables a simplified use of random walk simulations for the modelling of noise characteristics of nanoparticles attached to an electrode surface or during nanoimpact experiments. Here, it offers a simple and computationally efficient method to transform spherical objects into cubic geometries that are often significantly easier to implement.

\section{Acknowledgements}
The research leading to these results has received partial funding from the European Research Council under the European Union's Seventh Framework Programme (FP/2007-2013) / ERC Grand Agreement n. [320403]. Edward O. Barnes was supported by the Engineering and Physical Sciences Research Council (EPSRC) and St John's College, Oxford. Kay J. Krause and Bernhard Wolfrum gratefully acknowledge funding by the Helmholtz Young Investigator program.

\bibliographystyle{model1a-num-names}
\biboptions{sort&compress}

\end{document}